\documentclass[12pt,a4paper]{article}
\usepackage{amsmath}
\usepackage{amsthm}
\usepackage{amssymb}
\newcommand{\ds}{\displaystyle}
\newcommand{\p}{\partial}
\renewcommand{\deg}{\mathrm{deg}\,}
\newcommand{\ord}{\mathrm{ord}\,}
\newtheorem{thm}{Theorem}

\newtheorem{pro}{Proposition}

\title{\large\bf Symmetries and conservation laws for the Karczewska--Rozmej--Rutkowski--Infeld equation}
\author{A. Sergyeyev$^a$ and R. Vitolo$^b$\\[3mm]
$^a$~Mathematical Institute, Silesian University in Opava,\\ Na
Rybn\'\i{}\v{c}ku 1, 746 01 Opava,~Czech Republic\\
\texttt{Artur.Sergyeyev@math.slu.cz}
\\[2mm]
$^b$~Department of Mathematics and Physics ``E. De Giorgi'',\\ University of Salento, Lecce,
Italy
\\
\texttt{raffaele.vitolo@unisalento.it}
}
\begin{document}
\maketitle
\begin{abstract}\protect\vspace*{-5mm}
  We give a complete description of generalized symmetries and
  local conservation laws for the fifth-order Karczewska--Rozmej--Rutkowski--Infeld equation describing
  shallow water waves in a channel with variable depth. In particular, we show that
  this equation has no genuinely generalized symmetries and thus is not symmetry integrable.\looseness=-1
\end{abstract}
% \maketitle

\section*{Introduction}

Investigation of dynamics of shallow water waves has a long and distinguished history 
and remains a subject of intense research nowadays, see e.g.\
\cite{bus, c, bis} and references therein. Recently the authors of \cite{bus} have developed a systematic procedure for
deriving an equation for surface elevation of shallow water waves for a
prescribed relation between the orders of the two expansion parameters. This
procedure was {\em mutatis mutandis} applied in \cite{kri, krr}
for deriving a fifth-order equation describing unidirectional shallow-water waves in channels
with variable bottom geometry.

Consider the equation in question \cite{kri, krr}, to which we shall
refer to as to the Karczewska--Rozmej--Rutkowski--Infeld (KRRI)
equation,
%using a modification of the approach of \cite{bus}:
\begin{equation}
  \label{kri}
%\hspace*{-3mm}
\begin{array}{rcl}
    u_t&=&\displaystyle-u_x-\frac32 a u u_x-\frac16 b u_{xxx}
    +\frac38 a^2 u^2 u_x-\frac{23}{24}a b u_x u_{xx}
    -\frac{5}{12}a b u u_{xxx}\\[3mm]
    &&\hspace*{-13mm}\displaystyle+\frac{d}{2}(h'u+ h u_x)
    +\frac14 b d (-h'''u- h'' u_x+ h' u_{xx}+h u_{xxx})
    -\frac{19}{360}b^2 u_{xxxxx},
  \end{array}
\end{equation}
where $a$, $b$, and $d$ are constants (denoted in \cite{kri, krr} by
$\alpha,\beta$ and $\delta$), and $h=h(x)$ is a smooth function of
$x$ giving the dimensionless channel depth,
and the primes indicate $x$-derivatives of $h$; the dependent variable
$u$, i.e., the dimensionless wave elevation, denoted in \cite{kri, krr}
by $\eta$, is a function of $x$ and $t$.

For $a=0$ equation (\ref{kri}) becomes linear, while if $b=0$ we
obtain a first-order equation
\[
u_t=\displaystyle-u_x-\frac32 a u u_x+\frac38 a^2 u^2 u_x+\frac{d}{2}(h'u+ h
u_x),
\]
so both of these special cases of (\ref{kri}) are obviously integrable.

On the other hand, the authors of \cite{kri} have conjectured that
in general the KRRI equation (\ref{kri}) is not integrable by the
inverse scattering transform. We provide strong evidence to support
this conjecture by rigorously proving that (\ref{kri}) is not
symmetry integrable (see e.g.\ \cite{ms} and references therein 
for details on symmetry integrability), 
i.e., it has only finitely many generalized
symmetries, all of which are equivalent to the Lie point ones, and
just one local conservation law (\ref{lcl}). This also lends
substantial support to the conjectured absence of (non-dissipating)
multisoliton solutions for the equation under study, as well as of 
the zero-curvature representation involving a nonremovable parameter (as for the latter, 
cf.\ e.g.\ \cite{f, ma} and references therein). The proofs make
substantial use of the formal series technique, see e.g.\ \cite{ms}
and \cite{o} and references therein; cf.\ also \cite{v}.

\section{Main results}
\begin{thm}\label{t}
  The KRRI equation (\ref{kri}) with $a\neq 0$, $b\neq 0$ and $d\neq 0$ has no generalized
  symmetries of order greater than 9 and no local cosymmetries
  of order greater than 10.\looseness=-1
\end{thm}

Using this result we can readily compute {\em all} generalized symmetries,
cosymmetries and conservation laws for (\ref{kri}), as it suffices to find
all symmetries up to order 10 and all cosymmetries up to order 9.
%This yields the following assertions.
\begin{pro}\label{cs}
  If $a\neq 0$, $b\neq 0$, and $d\neq 0$, then the KRRI equation (\ref{kri}) has just
  one local conservation law
  \begin{equation}\label{lcl}
  \begin{array}{rcl}
    u_t&=&\displaystyle\left(u-\frac34 a u^2-\frac16 b u_{xx}+\frac18 a^2 u^3
    -\frac{13}{48}a b u_x^2-\frac{5}{12}a b u u_{xx}+\frac{d}{2} h u\right.
    \\[5mm]
   &&\left.+\displaystyle\frac14 b d (-h''u+ h u_{xx})-\frac{19}{360}b^2 u_{xxxx}\right)_x
   \end{array}
  \end{equation}
  with the density $u$ associated to the only cosymmetry $\gamma=1$ of
  (\ref{kri}).
\end{pro}

The result of Proposition~\ref{cs} can be seen as an ultimate amplification
of the results %on conservation laws
of \cite{kri2} on conservation laws of (\ref{kri}) for constant $h$.

As an aside note that since the density of the conservation law (\ref{lcl}) is just $u$, equation
(\ref{kri}) is in normal form with respect to low-order conservation laws in
the sense of \cite{ps}.

Denote by $F$ the right-hand side of (\ref{kri}).

\begin{pro}\label{s}
  If $a\neq 0$, $b\neq 0$, and $d\neq 0$ then all generalized symmetries of
  the KRRI equation (\ref{kri}) are equivalent to the Lie point ones.

  If $a\neq 0$, $b\neq 0$, $d\neq 0$, and $h'\neq 0$, then the only generalized
  symmetry of (\ref{kri}) is the one with the characteristics equal to $F$;
  this corresponds to the Lie point symmetry
  $\partial/\partial t$, i.e., the time translation.

  If $a\neq 0$, $b\neq 0$, $d\neq 0$ and $h=\mathrm{const}$, then, in addition to the time translation, we have a
  symmetry with the characteristics $u_x$, which corresponds to the
  Lie point symmetry $\partial/\partial x$, i.e., the space translation.

  Moreover, if $a\neq 0$, $b\neq 0$, $d\neq 0$, $h=\mathrm{const}$, and 
  $h d=4$, then in addition to the space and time translations equation
  (\ref{kri}) admits a symmetry with  the
  characteristics $5 t F+(x+2t)u_x+2 u-4/a$, which corresponds to a
  Lie point symmetry $5 t\partial/\partial t+(x+2t)\partial/\partial
  x+(4/a-2u)\partial/\partial u$.

\end{pro}

One of the immediate consequences of the above result is that (\ref{kri}) is not Lie remarkable
in the sense of \cite{m}, i.e., it is not uniquely determined by its Lie point symmetries.
Note also that it could be of interest to explore the point equivalence
transformations for (\ref{kri}), cf.\ e.g.\ \cite{vps} and references therein.

\section{Preliminaries}\label{pre}

In this section we recall a number of known definitions and results
from the so-called formal symmetry approach to integrability mostly
following \cite{ms,o}; cf.\ also \cite{g, kv}.

As usual, denote by $F$ the right-hand side of (\ref{kri}), by $u_i$
the $i$th $x$-derivative of $u$ (so $u_0\equiv u$, $u_1\equiv u_x$,
etc.), and by $D_x$ and $D_t$ total $x$- and $t$-derivatives
restricted to the differential equation (\ref{kri}) and its
differential consequences, i.e.,
%\begin{align}
\[
  D_x = \frac{\partial}{\partial x}
  +\sum\limits_{j=0}^\infty u_{j+1}\frac{\partial}{\partial
  u_j},\qquad
  %\\
  D_t = \frac{\partial}{\partial t} + \sum\limits_{j=0}^\infty D_x^j(F)
  \frac{\partial}{\partial u_j}.
%\end{align}
\]

For any {\em local} function, i.e., a smooth function
$f=f(t,x,u_0,u_1,\ldots)$ which may depend on $x,t$ and at most
finitely many $u_j$, define its order $\ord f$ as the greatest
integer $k$ such that $\p f/\p u_k\neq 0$, and if $f=f(x,t)$ we set
$\ord f=-\infty$ by definition. We denote by $\mathcal{A}$ the
algebra of local functions with respect to the usual multiplication.

Consider a local {\em conserved vector} for (\ref{kri}), cf.\ e.g.\
\cite{ps} and references therein, i.e., a pair of local functions
$(\rho, \sigma)$, that is, the density $\rho$ and the flux $\sigma$,
which satisfy the equation
\[
D_t(\rho)=D_x(\sigma).
\]
Two local conserved vectors $(\rho,\sigma)$ and
$(\tilde\rho,\tilde\sigma)$ are said to be equivalent if there
exists a function $\zeta\in\mathcal{A}$ such that
$\rho=\tilde{\rho}-D_x(\zeta)$ and
$\sigma=\tilde{\sigma}-D_t(\zeta)$. An equivalence class of
conserved vectors is called a local {\em conservation law} for
(\ref{kri}).\looseness=-1

Next, consider an algebra $\mathcal{B}$ of formal series of the form
\begin{equation}\label{l}
  L=\sum\limits_{i=-\infty}^k a_i D^i
\end{equation}
where the coefficients $a_i$ belong to $\mathcal{A}$ and the
indeterminate $D$ can be informally seen as just another avatar of
$D_x$.

The multiplication defined on monomials by the generalized Leibniz
rule
\[
a D^i\circ b D^j=a\sum\limits_{k=0}^\infty
\frac{i(i-1)\cdots(i-k+1)}{k!}D_x^{k}(b)D^{i+j-k}
\]
extends by linearity to the whole $\mathcal{B}$. It is readily seen
that this multiplication is associative and hence the commutator
$[A,B]=A\circ B-B\circ A$ turns $\mathcal{B}$ into a Lie algebra. We
shall omit $\circ$ whenever this does not lead to confusion.

The action of total derivatives extends from $\mathcal{A}$ to
$\mathcal{B}$ in an obvious manner,
\[
D_t\left(\sum\limits_{i=-\infty}^k a_i
D^i\right)=\sum\limits_{i=-\infty}^k D_t(a_i) D^i,
\]
and likewise for $D_x$.

As usual, the greatest $k$ such that $a_k\neq 0$ in (\ref{l}) is
called the degree of $L$ and denoted by $\deg L$, with the
convention that $\deg 0=-\infty$.

An $L\in\mathcal{B}$ with $\deg L\neq 0$ is called a {\em formal
recursion operator} (or {\em a formal symmetry}) of (\ref{kri}) of
rank $k$ if we have
\[
\deg (D_t(L)-[F_*,L])\leq \deg L+\deg F_*-k,
\]
where
%$F$ denotes the right-hand side of (\ref{kri}) and
for any
$f\in\mathcal{A}$ we define
\[
f_*=\sum\limits_{i=0}^{\ord f}\frac{\p f}{\p u_i}D^i.
\]
%where the sum actually always includes only finitely many terms.

Likewise, an $L\in\mathcal{B}$ is called {\em a formal symplectic
operator} (or {\em a formal conservation law}) for (\ref{kri}) of
rank $k$ if we have
\begin{equation}\label{fcl}
  \deg (D_t(L)+F_*^\dagger\circ L+L \circ F_*)\leq \deg L+\deg
  F_*-k.
\end{equation}
Here for any
\[
Q=\sum\limits_{j=-\infty}^q b_j D^j
\]
from $\mathcal{B}$ the formal adjoint $Q^\dagger$ is defined as
\[
Q^\dagger=\sum\limits_{j=-\infty}^q (-D)^j \circ b_j.
\]

In this connection recall (see e.g.\ \cite{ms,o,g} and references therein) 
that the characteristics $G$ of
generalized symmetries of (\ref{kri}) are local solutions of the
equation
\begin{equation}\label{sym}
  D_t(G)=\ell_F (G),
\end{equation}
where $\ell_F=F_*|_{D=D_x}$, while local cosymmetries are identified
with local solutions $\gamma$ of the equation
\begin{equation}\label{cosym}
  D_t(\gamma)=-\ell_F^\dagger (\gamma),
\end{equation}
where $\ell_F^\dagger=F_*^\dagger|_{D=D_x}$.

It is well known that if $(\rho,\sigma)$ is a local conserved vector for
(\ref{kri}) then $\delta\rho/\delta u$ is a cosymmetry for
(\ref{kri}). Here the Euler operator $\delta/\delta u$ is defined as
\[
\displaystyle\frac{\delta}{\delta u}=\sum\limits_{i=0}^\infty
(-D_x)^i \frac{\partial}{\partial u_i}.
\]
%is the Euler operator.

As the image of $\mathcal{A}$ with respect to $D_x$ lies in the
kernel of $\delta/\delta u$, the cosymmetry $\delta\rho/\delta u$
actually corresponds to a conservation law rather than to a
particular conserved vector. On the other hand, it is in general
{\em not} true that to any given cosymmetry there corresponds a
conservation law, cf.\ e.g.\ \cite{kv, o, ps} and references therein for
details.\looseness=-1

We identify the order of symmetry with that of its characteristics, and the
order of cosymmetry is handled in a similar manner.

In closing note that (see e.g.\ \cite{mag,o}) a generalized symmetry
of (\ref{kri}) is equivalent to a Lie point one if and only if the
characteristics $G$ of the symmetry in question can be written in
the form
\[
G=c(t)F+g_1(x,t,u)u_x+g_0(x,t,u)
\]
for suitable smooth functions $c,g_1,g_0$. The associated Lie point
symmetry then reads, up to the sign,
\[
c(t)\displaystyle\frac{\partial}{\partial
t}+g_1(x,t,u)\frac{\partial}{\partial
x}-g_0(x,t,u)\frac{\partial}{\partial u}.
\]

\section{Formal recursion and symplectic operators for the KRRI equation} %of Theorem~\ref{t}}
Theorem~\ref{t} is actually an immediate corollary of two stronger results:
\begin{thm}\label{t1a}
  If $a\neq 0$, $b\neq 0$, and $d\neq 0$, then the KRRI equation (\ref{kri}) has no formal
  recursion operator of rank greater than 9.
\end{thm}
% Then by the well known results \cite{mik1,mik2,mik3,sok,olv} we immediately
% obtain
% \begin{cor}\label{c1}
%   If $a\neq 0$ then eq.(\ref{kri}) has no generalized symmetries of order
%   greater than 8.
% \end{cor}
\begin{thm}\label{t2a}
  If $a\neq 0$, $b\neq 0$, and $d\neq 0$, then the KRRI equation (\ref{kri}) has no formal
  symplectic operator of rank greater than 10.
\end{thm}

Indeed, it is well known, see e.g.\ \cite{ms,o} and references
therein, that if a (1+1)-dimensional scalar evolution equation
admits no formal recursion operator of nonzero degree of rank $k$ or
greater then it cannot have generalized symmetries of order $k$ or
greater. Likewise, if a (1+1)-dimensional scalar evolution equation
admits no formal  symplectic operator of rank $k$ or greater then it
cannot have local cosymmetries of order $k$ or greater.

On the other hand, finding generalized symmetries (resp.\ local cosymmetries) up to a given order $k$ is just a matter of straightforward albeit somewhat tedious computation.\looseness=-1

Finally, as to any local conservation law of (\ref{kri}) there corresponds
a local cosymmetry of (\ref{kri}), finding all local conservation laws for
(\ref{kri}) when all local cosymmetries for (\ref{kri}) are known
becomes a straightforward matter too.

Let us also note that the absence of formal symplectic operator 
of rank 10 or higher (and hence {\em a fortiori} of infinite rank)
implies that the KRRI equation (\ref{kri}) admits no Hamiltonian or symplectic structure that 
can be written as a formal power series in the total $x$-derivative 
with local coefficients, cf.\ Section~\ref{pre}. While in principle 
it could happen that the KRRI equation admits some very exotic Hamiltonian or symplectic structure
involving complicated nonlocalities, cf.\ e.g.\ \cite{s05} and references therein, this is extremely unlikely.\looseness=-1  
 
\subsection{Proof of Theorem~\ref{t1a}}
Seeking a contradiction, suppose that there exists an $L\in\mathcal{B}$ with $\deg L\neq 0$
such that
\begin{equation}\label{fs0}
\deg (D_t(L)-[F_*,L])\leq \deg L+\deg F_*-10.
\end{equation}

Without loss of generality we can assume (cf.\ e.g.\ \cite{ms,o})
that our formal recursion operator $L$ has $\deg L=1$, and let
\[
L=f D+\sum\limits_{j=0}^\infty s_j D^{-j},
\]
where $f\in\mathcal{A}$ and $s_j\in\mathcal{A}$,
so (\ref{fs0}) boils down to
\begin{equation}\label{fs}
  \deg (D_t(L)-[F_*,L])\leq -4.
\end{equation}
Thus, we need to equate to zero the coefficients at $D^j$ for
$j=5,4,\dots,-3$ (the coefficients at the higher powers of $D$
vanish automatically) in
\[
M=D_t(L)-[F_*,L].
\]

Equating to zero the coefficient at $D^5$ in $M$ yields
\[
b^2 D_x(f)=0,
\]
so, as $b\neq 0$ by assumption, we have $D_x(f)=0$ and hence
\begin{equation}\label{f}
  f=f(t),
\end{equation}
i.e., $f$ is an arbitrary smooth function of $t$ alone.

Next, equating to zero the coefficient at $D^4$ in $M$ while using (\ref{f})
yields
\[
b^2 D_x(s_0)=0,
\]
whence
\[
s_0=f_0(t),
\]
where $f_0$ again is an arbitrary smooth function of $t$.  As
$f_0(t)$ commutes with all elements of $\mathcal{B}$, without loss
of generality put $s_0=0$.

In a similar fashion as above, equating to zero the coefficients at $D^i$,
$i=3,2,\dots,-3$ in $M$ yields equations of the form
\begin{equation}\label{k0}
  D_x(s_j)=K_j,
\end{equation}
where $j=1,\dots,7$ and $K_j$ are some local functions.

As the image of $\mathcal{A}$ with respect to $D_x$ lies in the kernel of the
Euler operator
$\delta/\delta u$,
%\[
%\delta/\delta u=\sum\limits_{i=0}^\infty (-D_x)^i \partial/\partial u_i,
%\]
see e.g.\ \cite{ms,o} for details, and $s_j$ are local functions by
assumption, we have the necessary conditions for (\ref{k0}) to hold
of the form
\begin{equation}\label{k}
  \delta K_j/\delta u=0,\quad j=1,\dots,7.
\end{equation}

Using our assumptions that $a,b,d\neq 0$, $f$ depends on $t$ alone,
and $s_0=0$, straightforward but somewhat tedious computations show
that the only nontrivial conditions among (\ref{k}) are those for
$j=5$ and $j=7$.

Recursively solving equations (\ref{k0}) for $j=1,\dots,4$ we find that the
condition (\ref{k}) for $j=5$ reads
\[
\frac{\partial f}{\partial t}\frac{a}{b}=0.
\]
As $a\neq 0$ by assumption, we see that $f$ is actually a constant rather than
a function of $t$.

With this in mind we can readily solve (\ref{k0}) for $j=5,6$, and then we find
that differentiating the condition
\[
\delta K_7/\delta u=0,
\]
with respect to $u_{xx}$ and then to $u_x$ yields
\[
a^3 f/b=0.
\]
As $a\neq 0$ and $b\neq 0$ by assumption, we see that $f=0$, i.e.,
$\deg L<1$, which contradicts our initial assumption. Hence, a
formal recursion operator for (\ref{kri}) of rank greater than nine
and of nonzero degree does not exist, and the result follows.

\subsection{Proof of Theorem~\ref{t2a}}
Now, seeking a contradiction, just as in the proof of Theorem~\ref{t1a},
assume that
\[
L=\sum\limits_{j=-\infty}^r s_{j-r} D^j,
\]
where $s_j\in\mathcal{A}$, is a formal symplectic operator of rank $10$ for
(\ref{kri}).

Let
\[
N=D_t(L)+F_*^\dagger\circ L+L \circ F_*.
\]
The condition (\ref{fcl}) for $k=10$ is satisfied if and only if the
coefficients of $N$ at the powers $D^j$, $j=r+4,\dots, r-4$ vanish
(vanishing of the coefficients at $D^{r+5}$ and higher powers of $D$
occurs automatically).

In particular, vanishing of the coefficient at $D^{r+4}$ yields
\[
b^2 D_x(s_0)=0.
\]

Hence
\[
s_0=f_0(t),
\]
where again $f_0(t)$ is an
%we shall again denote by $f_j=f_j(t)$
arbitrary smooth function of $t$.

It is readily checked that vanishing of the coefficients at $D^j$ for
$j=r+3,\dots,r-4$ yields equations of familiar structure
(cf.\ the preceding subsection)
\begin{equation}\label{k0a}
  D_x(s_j)=H_j,
\end{equation}
where $j=1,\dots,8$ and $H_j$ are some local functions.

In complete analogy with the proof of Theorem~\ref{t1a}
%in the preceding section
we have the necessary conditions for solvability of (\ref{k0a}) of
the form
\begin{equation}\label{h}
  \delta H_j/\delta u=0,\quad j=1,\dots,8.
\end{equation}

Recursively solving (\ref{k0a}) with respect to $s_j$ we readily
find that the first nontrivial condition among (\ref{h}) is the one
for $j=4$, which reads
\[
a s_0 d \frac{\partial h}{\partial x}=0.
\]
We see that if $h\not\equiv \mathrm{const}$ then, as $a\neq 0$ and
$d\neq 0$ by assumption, $s_0=0$ and hence a formal symplectic
operator of rank 10 or higher for (\ref{kri}) does not
exist.\looseness=-1

Moreover, as the breakdown occurs at $j=4$, for $h\not\equiv \mathrm{const}$ we actually
have a much stronger result, namely, a formal symplectic operator of rank 6 or higher for
(\ref{kri}) in this case does not exist too.

Now turn to the case of $h\equiv \mathrm{const}$ when $s_0$ does not
have to vanish. Then we find that the next nontrivial condition
among (\ref{h}) is the one for $j=6$, and this condition reads
\[
a b \displaystyle\frac{\partial f_0}{\partial t}=0,
\]
so $s_0=f_0=\mathrm{const}$.

With this in mind we can readily solve recursively further equations
from (\ref{k0a}), and we see that the next nontrivial condition from
(\ref{h}) occurs for $j=8$.

In particular, we have
\[
\ds\frac{\partial^2 (\delta H_{8}/\delta u)}{\partial u_{xx} \partial u_x}=
\frac{27}{1444 b} a^3 s_0 (157 r+837).
\]

The expression on the right-hand side of this equation must vanish
because of (\ref{h}) for $j=8$. As $r$ is an integer and $a\neq 0$
by assumption, this is only possible if $s_0=0$, and hence $\deg
L<r$, which contradicts our initial assumption. Thus, a formal
symplectic operator of rank 10 or higher for (\ref{kri}) does not
exist even if $h\equiv \mathrm{const}$.

%%% \section{Proof of Proposition~\ref{cs}}

\section*{Acknowledgments}
This research was supported in part by the Ministry of Education,
Youth and Sports of the Czech Republic (M\v{S}MT \v{C}R) under RVO
funding for I\v{C}47813059, and by the Grant Agency of the Czech
Republic (GA \v{C}R) under grant P201/12/G028. The
computations in the paper were mostly performed using the computer algebra packages {\em
Jets} \cite{jets}, {\em CDE} \cite{cde} and CRACK \cite{w}.\looseness=-1

RV thanks Thomas Wolf for stimulating discussions and
gratefully acknowledges warm hospitality extended to him in the course of his visit
to Silesian University in Opava where the work on the present paper was initiated.\looseness=-1

\end{document}